\documentclass[preprint,aps]{revtex4-2}

\usepackage[utf8]{inputenc} 
\usepackage{graphicx}
\usepackage{bm}
\usepackage{nicefrac}
\usepackage{array} 
\usepackage{upgreek}
\usepackage{amsmath}
\usepackage{makecell} 
\usepackage{booktabs}
\usepackage{multirow} 
\usepackage{titlesec}
\usepackage{placeins} 
\usepackage{hyperref}
\usepackage{cleveref}
\usepackage{titlesec}
\usepackage{tabularx}

\titleformat{\section}
{\normalfont\Large\bfseries} 
{\thesection}{1em}{}         

\crefname{figure}{Fig.}{Figs.}
\crefname{equation}{Eq.}{Eqs.}

\begin{document}

\titleformat{\section}
  {\normalfont\large\bfseries}{\thesection}{1em}{}
\titleformat{\subsection}
  {\normalfont\normalsize\bfseries}{\thesubsection}{1em}{}
\titleformat{\subsubsection}
  {\normalfont\normalsize\itshape}{\thesubsubsection}{1em}{}

\renewcommand\thesection{\arabic{section}}
\renewcommand\thesubsection{\thesection.\arabic{subsection}}
\renewcommand\thesubsubsection{\thesubsection.\arabic{subsubsection}}

\title{Electronic Structure and Superconducting Gap of HgBa$_2$Ca$_2$Cu$_3$O$_{8+\delta}$ Revealed by Laser-Based Angle-Resolved Photoemission Spectroscopy}

\author{Taimin Miao$^{1,2\sharp}$, Wenshan Hong$^{1\sharp}$, Qinghong Wang$^{1,2\sharp}$, Shanshan Zhang$^{1,2\sharp}$, Bo Liang$^{1,2}$,  Wenpei Zhu$^{1,2}$, Neng Cai$^{1,2}$, Mingkai Xu$^{1,2}$, Shenjin Zhang$^{3}$, Fengfeng Zhang$^{3}$, Feng Yang$^{3}$, Zhimin Wang$^{3}$, Qinjun Peng$^{3}$, Zuyan Xu$^{3}$, Hanqing Mao$^{1}$, Zhihai Zhu$^{1}$, Xintong Li$^{1}$, Guodong Liu$^{1}$, Lin Zhao$^{1}$, Yuan Li$^{1}$ and X. J. Zhou$^{1,2,4*}$}

\affiliation{
\\$^{1}$Beijing National Laboratory for Condensed Matter Physics, Institute of Physics, Chinese Academy of Sciences, Beijing 100190, China
\\$^{2}$University of Chinese Academy of Sciences, Beijing 100049, China
\\$^{3}$Technical Institute of Physics and Chemistry, Chinese Academy of Sciences, Beijing, China
\\$^{4}$Songshan Lake Materials Laboratory, Dongguan 523808, China
\\$^{\sharp}$These people contributed equally to the present work.
\\$^{*}$Corresponding authors: xjzhou@iphy.ac.cn
}

\date{\today}

\maketitle

{\bf The spatially-resolved laser-based high resolution angle resolved photoemission spectroscopy (ARPES) measurements have been performed on the optimally-doped HgBa$_2$Ca$_2$Cu$_3$O$_{8+\delta}$ (Hg1223) superconductor with a \(T_c\) at 133\,K. Two distinct regions are identified on the cleaved surface: the single Fermi surface region where only one Fermi surface is observed, and the double Fermi surface region where two Fermi surface sheets are resolved coming from both the inner (IP) and outer (OP) CuO\(_2\) planes. The electronic structure and superconducting gap are measured  on both of these two regions. In both cases, the observed electronic states are mainly concentrated near the nodal region. The momentum dependence of superconducting gap deviates from the standard d-wave form. These results indicate that the surface electronic structure of Hg1223 behaves more like that of underdoped cuprates.}

\section*{Introduction}

The cuprate superconductors exhibit markedly different superconducting transition temperatures (\(T_c\)) in different classes of materials\cite{chu_2015_PhysicaC:SuperconductivityanditsApplications}. Uncovering the origin of this \(T_c\) difference is important for clarifying the mechanism of high temperature superconductivity. To achieve this goal, it is essential to understand the electronic structure and superconducting gap structure of the cuprate superconductors\cite{keimer_2015_Naturea}. Angle-resolved photoemission spectroscopy (ARPES) has proven particularly powerful in these efforts\cite{damascelli_2003_Rev.Mod.Phys.,schrieffer_2007_,sobota_2021_Rev.Mod.Phys.}. However, most ARPES studies have been focused on the Bi-based cuprates because they can provide clean and stable cleaved surface for ARPES to probe intrinsic electronic structures\cite{damascelli_2003_Rev.Mod.Phys.,schrieffer_2007_,sobota_2021_Rev.Mod.Phys.}. The Hg-based HgBa$_2$Ca$_2$Cu$_3$O$_{8+\delta}$ (Hg1223) superconductor exhibits the highest \(T_c\) of 134\,K at ambient pressure\cite{schilling_1993_Nature} and provides an ideal system to study high temperature superconductivity mechanism. However, APRES studies on Hg1223 superconductor remain remarkably scare\cite{horio_2025_Phys.Rev.Lett.}.

In the present paper, we report our investigations of the electronic structure and superconducting gap structure of Hg1223. By performing spatially-resolved laser-based ARPES, we observed two different regions on the cleaved surface. The Fermi surface, band structures and superconducting gap are measured on each of these two regions. 


\section*{Experimental Methods}
High-quality single crystals of Hg1223 were grown using  the self-flux method\cite{wang_2018_Phys.Rev.Mater.}. The crystals were post-annealed at 530\(^\circ\)C under an oxygen pressure of \(\sim\)12 atmospheres for 7 days, yielding optimally doped samples with a superconducting transition temperature \(T_c^{onset}\) of 133\,K and a narrow transition width of \(\sim\)2.5\,K, as shown in Fig. 1b. 
ARPES measurements were carried out using a bias laser-ARPES system equipped with the 6.994\,eV vacuum-ultra-violet (VUV) laser and a DA30L hemispherical electron energy analyzer\cite{liu_2008_Rev.Sci.Instrum.,taimin__RSI}. The copper sample holder was electrically connected to a Keithley 2450 source meter, allowing application of a well-controlled sample bias during measurements to enhance the covered momentum space\cite{taimin__RSI}. The laser beam size was focused to \(\sim\)10\,\(\mathrm{\mu} \)m on the sample surface. The overall energy resolution was \(\sim\)8.5\,meV with a 10\,eV pass energy and a 0.3\,mm slit at a -90\,V sample bias (Fig.~1d and 1e) and \(\sim\)4.5\,meV when using a 5\,eV pass energy and a 0.3\,mm slit at a -30\,V sample bias (Figs.~2-5). The angular resolution was better than 0.2\(^\circ\) with zero sample bias. The spatially resolved ARPES measurements were performed by scanning the sample position in real space to acquire the photocurrent map or band structure map.
Samples were cleaved in situ at a low temperature and measured at 30\,K under ultrahigh vacuum with a base pressure better than 2 \(\times\) 10\(^{-11}\) \,mbar. The Fermi level of the sample was referenced by measuring a clean polycrystalline gold.


\section*{Results and Discussion}


As shown in Fig.~1a, Hg1223 has a perfect tetragonal structure (space group \(P4/mmm\)) that contains a trilayer CuO\(_2\) block composed of two outer CuO\(_2\) planes (OP) and one inner CuO\(_2\) plane (IP) per unit cell, separated by Ca layers\cite{schilling_1993_Nature}. These trilayer blocks are isolated by BaO-HgO\(_\delta\)-BaO layers. It is generally believed that the excess interstitial oxygens, residing in the HgO\(_\delta\) layers, are responsible for doping the CuO\(_2\) planes. However, unlike Bi-based trilayer sample Bi\(_2\)Sr\(_2\)Ca\(_2\)Cu\(_3\)O\(_{10+\delta}\) (Bi2223)\cite{luo_2023_Nat.Phys.}, Hg1223 lacks a natural cleavage surface. As a result, obtaining a clean, well-defined, and stable surface has been a major challenge for ARPES measurements. There are four possible cleavage planes for Hg1223, as marked by dashed lines in Fig.~1a. Breaking between the HgO\(_\delta\)-BaO layers gives rise to two cleavage planes (1 and 4) while breaking between the OP CuO\(_2\) plane and BaO layer gives rise to cleavage planes 2 and 3. It was found that cleavage is easier to occur between the HgO\(_\delta\) and BaO layers, leading to the majority of surface terminations from the cleavage planes 1 and 4\cite{sreedhar_2020_Phys.Rev.B}. Moreover, since the HgO\(_\delta\) layers and the trilayer CuO\(_2\) blocks are polar, the polar terminations may induce charge redistribution near the surface after cleavage\cite{nakagawa_2006_Nat.Mater.,shuaishuailiLiShuaiShuai_2023_Chin.Phys.Ba}. This surface self-doping effect may complicate the effort to access the intrinsic bulk electronic structure using ARPES\cite{shuaishuailiLiShuaiShuai_2023_Chin.Phys.Ba}. 


We first carried out spatially resolved APRES measurements on Hg1223 by scanning the sample surface point-by-point and collecting the photocurrent or measuring the band structure at the same time. Fig.~1c shows the photocurrent map acquired over a large area of the cleaved surface. The intensity distribution exhibits clear spatial inhomogeneity. We mainly observed two kinds of regions on the surface with distinct Fermi surface topologies. For the region 1, only a single Fermi surface is observed (Fig.~1d) while two Fermi surface sheets are clearly resolved (Fig.~1e) for the region 2. The majority of the cleaved surface corresponds to Region 1, only a small fraction of the surface area belongs to Region 2.



Figure~2 presents a detailed Fermi surface and band structures of Hg1223 measured at 30\,K in Region~1 with the single Fermi surface. The Fermi surface map is shown in Fig.~2a and a constant energy contour at the binding energy of 30\,meV is shown in Fig.~2b. Fig.~2c displays the band structures measured along a series of momentum cuts indicated in Fig.~2b. A well-defined band is observed in the measured band structures near the nodal region (Cuts~1-6). As the momentum cuts move toward the antinodal region (Cuts~7-12), the lower binding energy part of the band becomes vague but the high binding energy part is still observable. The band becomes invisible when the momentum cuts further move toward the antinodal region (Cuts~13-14 and beyond). When the momentum cuts move from the nodal region to the antinodal region, the band top shift downwards to high binding energy, corresponding to the opening of the superconducting gap. 

The observed electronic states are mainly concentrated near the nodal region as seen in Fig.~2a and 2b. The Fermi momenta extracted from the band structures in Fig.~2c are plotted as red circles in Fig.~2a. The full Fermi surface topology can not be determined because of the lack of band structures observed near the antinodal region. Considering the observed Fermi surface segment as a part of a large Fermi surface, we fit the extracted Fermi surface with a tight-binding model\cite{markiewicz_2005_Phys.Rev.B}: \(E(k)=\mu-2t(cos(k_xa)+cos(k_ya))-4t'cos(k_xa)cos(k_ya)-2t''(cos(2k_xa)+cos(2k_ya))\) which yields \(t'/t = -0.25\), \(t''/t = 0.22\) and \(\mu/t = -0.66\). 
The resulted tight-binding Fermi surface corresponds to a  hole doping level of \(\sim 0.07\).






Now we turn to the determination of the superconducting gap of Hg1223 for region 1 with the single Fermi surface. Fig.~3a shows the energy distribution curves (EDCs) extracted from Fig.~2c along the measured Fermi surface. The corresponding location of the Fermi momenta (P1-P15) indicated in Fig.~3c. A weak peak is observed in the EDCs only very closed to the nodal region. Fig.~3b shows the 
symmetrized EDCs obtained from Fig.~3a. The superconducting gap at each Fermi momentum can be extracted from the peak position of these symmetrized EDCs, as marked by vertical ticks in Fig.~3b. The extracted gap size is plotted in Fig.~3d as a function of the Fermi surface angle, \(\theta\).  The gap near the nodal region is fitted by a \(d\)-wave form, \(\Delta(\theta) = \Delta_0 \cos(2\theta)\), yielding \(\Delta_0\sim\)41\,meV. In Fig.~3e, we plot the same gap  against the \(d\)-wave form factor\(|\cos(k_x a)-\cos(k_y a)|/2\). The gap near the nodal region is fitted by a linear line, \(\Delta=\Delta_0\cdot |\cos(k_x a)-\cos(k_y a)|/2\), with \(\Delta_0\sim\)37\,meV. The gap size clear deviates from the d-wave form when the momentum point is slightly away from the nodal region. The momentum dependence of the superconducting gap is similar to that observed in underdoped cuprate superconductors\cite{hashimoto_2014_NaturePhys}.


Now we switch to region 2 where two Fermi surface sheets are observed. Fig.~4a shows the Fermi surface map measured at 30\,K and a constant energy contour at the binding energy of 30 meV is shown in Fig. 4b. Two closely spaced Fermi surface sheets are clearly visible. Fig. 4c displays the band structures measured along a series of momentum cuts indicated
in Fig. 4b. Two well-defined bands are observed in the measured band structures near the nodal region (Cuts 1-5), labeled as IP and OP which originate from the inner and outer CuO\(_2\) planes, respectively. As the momentum cuts move toward the antinodal region (Cuts 6-10), the bands near the Fermi level get weaker. The bands become invisible when the momentum cuts further move toward the antinodal region (Cuts 11-12 and beyond). When the momentum cuts move from the nodal region to the antinodal region, the overall top of the two bands shift downwards to high binding energy, corresponding to the opening of the superconducting gap.

The observed electronic states are also mainly concentrated near the nodal region as seen in Fig.~4a and 4b. The extracted Fermi momenta of the IP and OP bands from Fig.~4c are plotted as red and blue circles in Fig.~4a, respectively. The full Fermi surface topology can not be determined because of the lack of band structures observed near the antinodal region. Considering the observed Fermi surface segments as a part of large Fermi surfaces, we fit the extracted IP and OP Fermi surfaces with the tight-binding model\cite{markiewicz_2005_Phys.Rev.B}, which  yields \(t'/t = -0.35\), \(t''/t = 0.24\), and \(\mu_{OP}/t = -0.62\) for the OP Fermi surface, and \(t'/t = -0.38\), \(t''/t = 0.26\), and \(\mu_{IP}/t = -0.16\) for the IP Fermi surface. The resulted tight-binding Fermi surfaces correspond to a hole doping level of \(\sim0.02\) for the OP Fermi surface and  \( \sim-0.1\) for the IP Fermi surface. The negative sign corresponds to electron doping and it is hard to believe that the inner CuO\(_2\) plane in Hg1223 is electron doped. Further work need to be done to get the full Fermi surface topology and understand their corresponding doping levels.

To determine the superconducting gap along the IP and OP Fermi surface sheets in region 2 of Hg1223, Fig.~5a and 5c show the EDCs along the IP and OP Fermi surface, respectively, extracted from Fig.~4c.  The corresponding Fermi momentum location of the IP (IP1-IP14) and OP (IP1-IP14) Fermi surface are indicated in Fig. 5e. Sharp coherent peaks are observed in the EDCs along the IP Fermi surface (Fig.~5a). Fig. 5b and Fig.5d shows the corresponding symmetrized EDCs obtained from Fig.~5a and Fig.~5c. The superconducting gap can be extracted from the peak position of the symmetrized EDCs, as marked by vertical ticks in Fig. 5b and 5d. The extracted gap size is plotted in Fig.~5f as a function of the Fermi surface angle, \(\theta\) for the IP (red circles) and OP (blue circles) Fermi surface. The gap is fitted by a \(d\)-wave form, \(\Delta(\theta) = \Delta_0 \cos(2\theta)\), yielding \(\Delta_0\sim52\)\,meV for IP and  \(\sim41\)\,meV for OP. In Fig.~5g, we plot the same gap against the \(d\)-wave form factor \(|\cos(k_x a)-\cos(k_y a)|/2\). The gap is fitted by a linear line, \(\Delta=\Delta_0\cdot |\cos(k_x a)-\cos(k_y a)|/2\), with \(\Delta_0\sim47\)\,meV for IP and \(\sim 35\)\,meV for OP. The IP Fermi surface exhibits a larger gap than the OP Fermi surface demonstrating a distinct layer dependent pairing strength in the double Fermi surface region.



The two regions with distinct Fermi surface topologies may come from different cleavage planes. Since the majority of the measured regions correspond to region 1 with single Fermi surface, and breaking the HgO\(\delta\)-BaO layers is easier than breaking between the OP CuO\(_2\) plane and BaO layer, it is likely that region 1 with single Fermi surface comes from the cleavage planes 1 and/or 4 (Fig.~1a). The region 2 with double Fermi surface may come from the cleavage planes 2 and/or 3 (Fig.~1a). In principle, two sets of Fermi surface should be observed in trilayer cuprates because they contain two types of CuO\(_2\) planes (IP and OP) per unit cell. The observation of two Fermi surface sheets in region 2 is consistent with such expectation. The absence of one set of Fermi surface in region 1 may be related to the photoemission matrix element effects associated with the particular cleavage plane(s). A Fermi surface comparison between the two regions suggests that the single Fermi surface in region 1 is similar to the OP Fermi surface in region 2. This indicates that the observed single Fermi surface in region 1 originates from the outer CuO\(_2\) plane, while the Fermi surface from the inner CuO\(_2\) plane is suppressed.

For the optimally doped Hg1223 sample, the observed results are unusual in several aspects. First, only a small portion of the Fermi surface near the nodal region is observed in both region 1 and region 2, with a rapid loss of spectral weight away from the nodal toward the antinodal regions. Such a narrow confinement of electronic states near the nodal region is usually observed in underdoped cuprate superconductors\cite{shen_2005_Science,peng_2013_NatCommuna}. Second, the extracted Fermi surface in both regions corresponds to a very low hole doping level, which is inconsistent with the optimally doped nature of the bulk crystal. Third, the momentum dependence of the superconducting gap in both regions deviates from a standard d-wave form, which is consistent with the behavior of the  underdoped cuprates. Overall, although the bulk crystal is optimally doped, our ARPES measurements suggest that the surface electronic structure in both regions behaves more like that of underdoped cuprates. These effects could be influenced by the polar nature of the cleavage surface, variations in local termination, or even subtle oxygen loss during cleaving, any of which may modify the near-surface doping environment. Further measurements under more controlled surface conditions will likely be needed to clarify these possibilities.

\section*{Summary}
In summary, we have performed spatially-resolved laser-based ARPES measurements on the Hg1223 superconductor. We have identified two distinct regions on the cleaved surface: the majority region where only one Fermi surface is observed, and the minority region where double Fermi surface sheets are observed. The electronic structure and superconducting gap are measured  on both of these two regions. The observed electronic states are mainly concentrated near the nodal region in both cases. The momentum dependence of superconducting gap deviates from the standard d-wave form. These results suggest that the surface electronic structure in both regions behaves more like that of underdoped cuprates. Further work need to be done to control the doping level and investigate the doping evolution of the electronic structure and superconducting gap of Hg1223.



\newpage


\noindent {\bf Acknowledgement}\\
This work is supported by the National Key Research and Development Program of China (Grant Nos. 2021YFA1401800, 2022YFA1604200, 2022YFA1403900, 2023YFA1406002, 2024YFA1408301 and 2024YFA1408100), the National Natural Science Foundation of China (Grant Nos. 12488201, 12374066, 12374154 and 12494593),  Quantum Science and Technology-National Science and Technology Major Project (Grant No. 2021ZD0301800), CAS Superconducting Research Project (Grant No. SCZX-0101) and the Synergetic Extreme Condition User Facility (SECUF).\\

\noindent {\bf Author Contributions}\\
T.M.M. and X.J.Z. proposed and designed the research. T.M.M. carried out the ARPES experiments.  W.S.H.,S.S.L. and Y.L. contributed to the crystal growth. T.M.M. and Q.H.W. contributed to the sample preparation. T.M.M., B.L., W.P.Z., N.C, M.K.X., S.J.Z., F.F.Z., F.Y., Z.M.W., Q.J.P., Z.Y.X., M.H.Q., Z.H.Z., X.T.L.,   G.D.L., L.Z. and X.J.Z. contributed to the development and maintenance of Laser ARPES systems. T.M.M. and X.J.Z. analyzed the data and wrote the paper. All authors participated in discussions and comments on the paper.


\newpage

\begin{figure}[tbp]
  \begin{center}
  \includegraphics[width=1\columnwidth,angle=0]{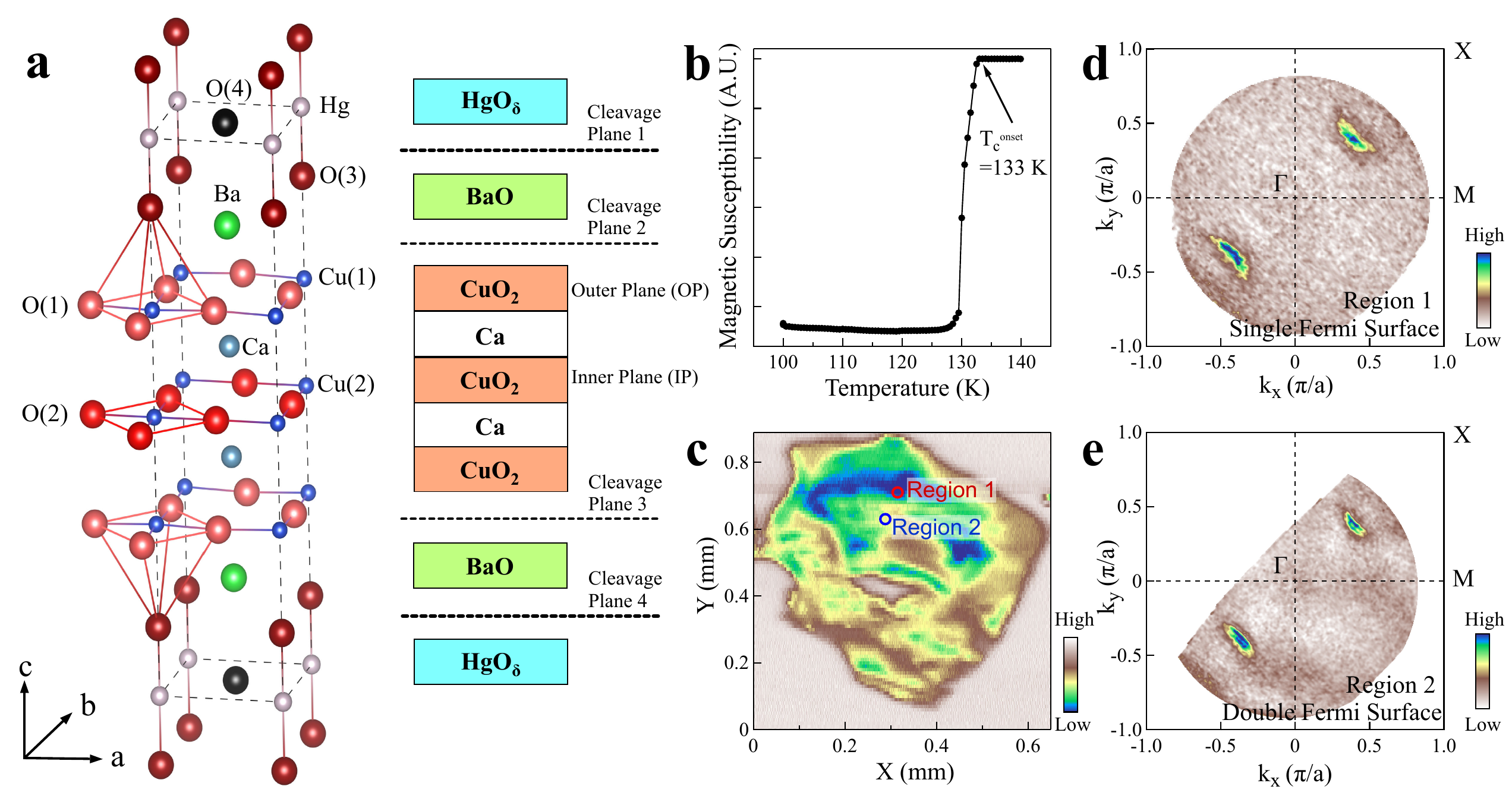}
  \end{center}
  \caption{\textbf{Spatially resolved ARPES measurements on different cleavage surfaces of Hg1223.}  (a) Crystal structure of the trilayer cuprate Hg1223, consisting of three CuO\(_2\) planes per unit cell. The inner CuO\(_2\) plane (IP) is sandwiched between two outer planes (OPs) separated by Ca layers. The right panel illustrates several possible cleavage planes (HgO\(_\delta\), BaO or CuO\(_2\)).  (b) AC magnetic susceptibility measurement under a 10\,Oe field showing a sharp superconducting transition at \(T_c=133\)\,K with a transition width (10\%-90\%) of 2.5\,K. (c) Real space photocurrent map acquired by point-by-point scanning the sample surface. Two kinds of regions with distinct Fermi surface topology are observed on the surface. (d) Fermi surface mapping obtained from Region 1 (red circle in c) where only a single Fermi surface near the nodal region is observed. To enhance the covered momentum space, we measurement was carried out by applying a bias voltage of -90\,V on the sample. (e) Fermi surface mapping obtained from Region 2 (blue circle in c) where two Fermi surface sheets near the nodal region are observed. 
  The measurement was also performed under a -90\,V sample bias.
  }
  \label{Fig1_crystal}
\end{figure}

\begin{figure}[tbp]
  \begin{center}
  \includegraphics[width=1\columnwidth,angle=0]{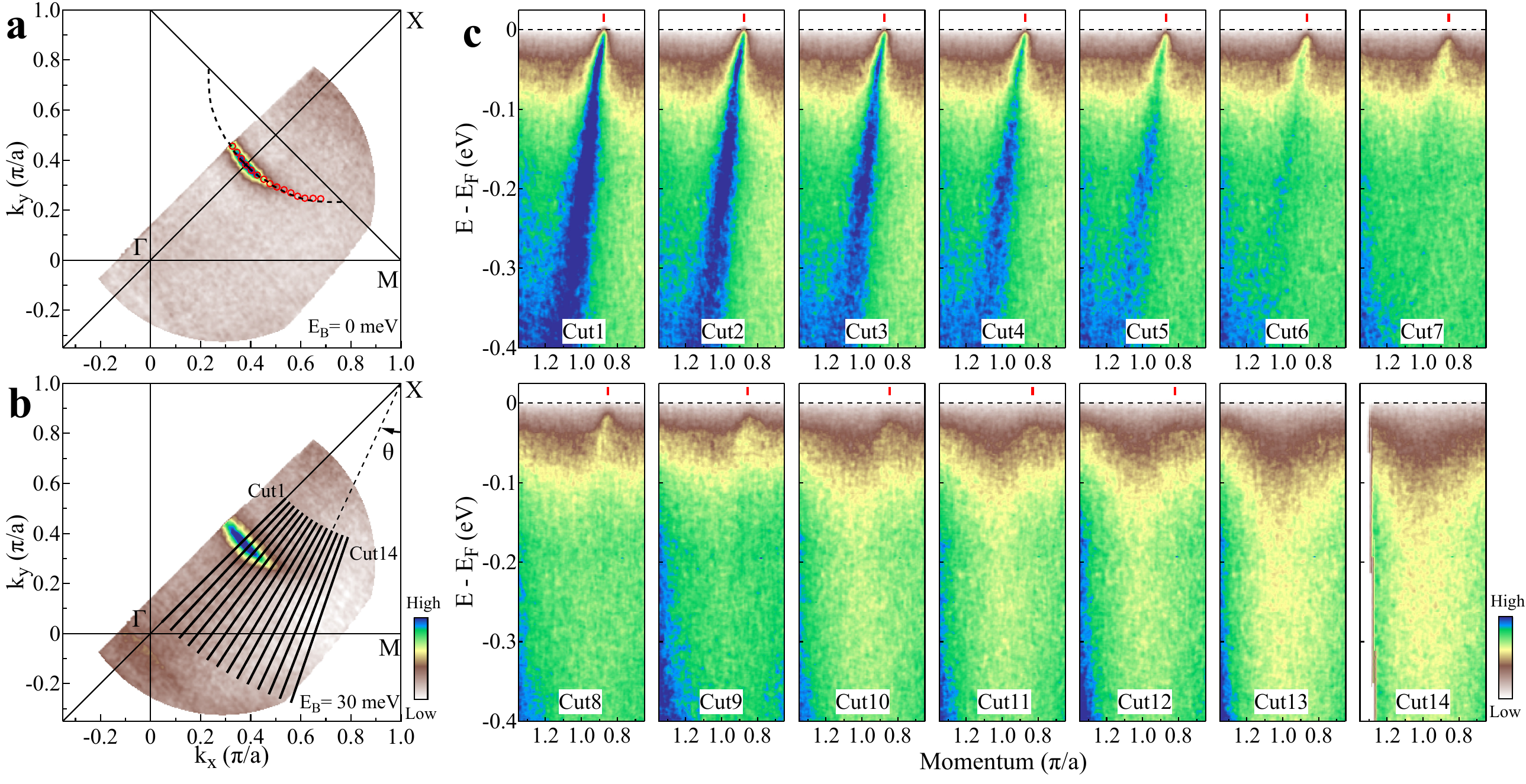}
  \end{center}
  \caption{\textbf{Electronic structure of Hg1223 measured at 30\,K from the single Fermi surface region.} The measurement was performed under a sample bias of -30\,V. (a) Fermi surface mapping of Hg1223 measured at 30\,K in the superconducting state. The red open circles denote the experimentally determined Fermi momenta (\(k_F\)), and the black dashed curve represents guide to the eyes. (b) The corresponding constant energy contour at a binding energy of 30\,meV. (c) Band structures measured along different momentum cuts. The location of the momentum cuts is shown by black lines in (b). All the momentum cuts cross the \((\pi,\pi)\) point when extrapolated. The vertical tick in each panel mark the determined Fermi momenta.}
  \label{Fig2_SingleBandBandStru}
\end{figure}

\begin{figure}[tbp]
    \begin{center}
    \includegraphics[width=0.8\columnwidth,angle=0]{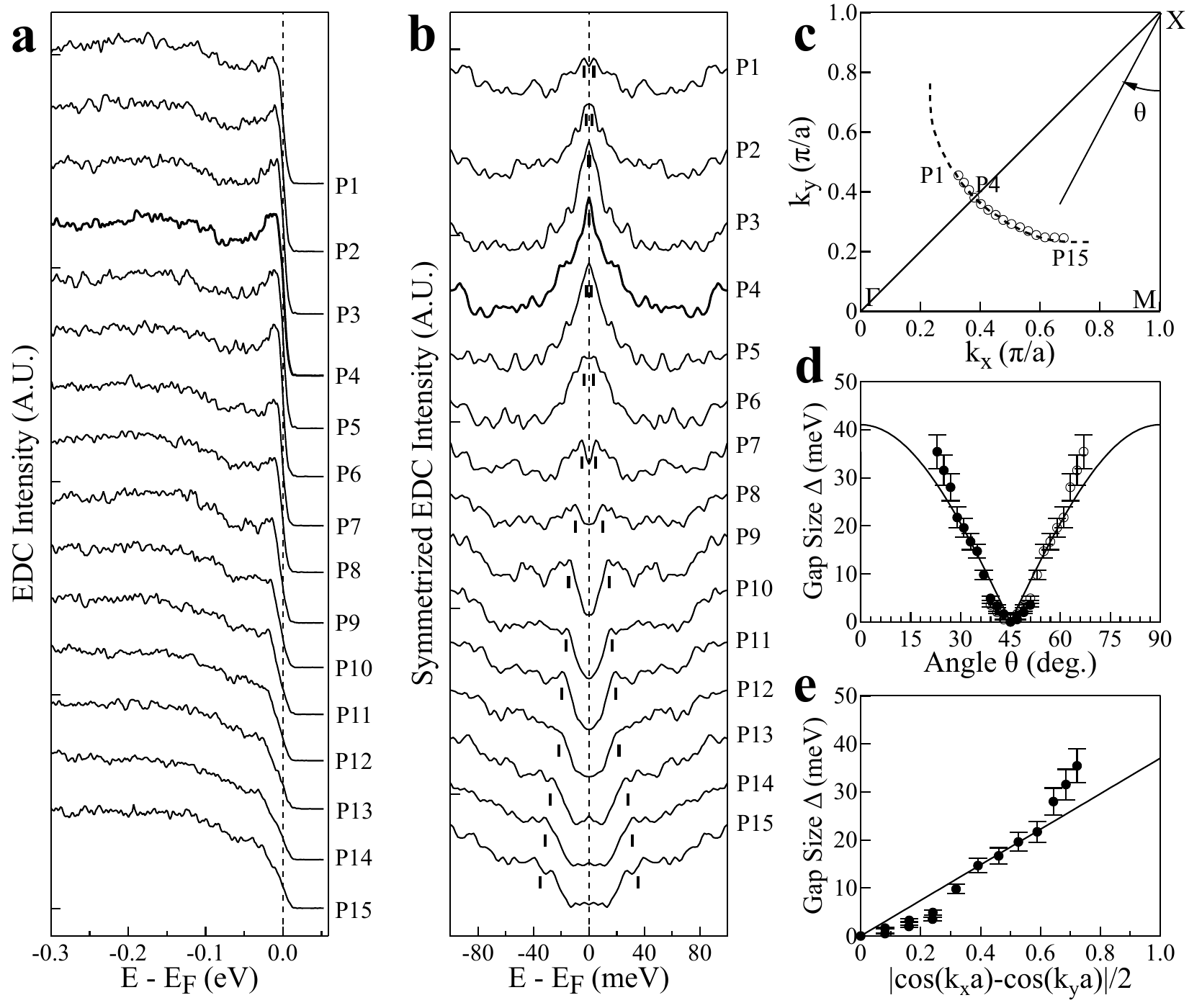}
    \end{center}
   \caption{\textbf{Superconducting gap of Hg1223 measured in the single Fermi surface region at 30\,K.} (a) Energy distribution curves (EDCs) measured along the Fermi surface of the single Fermi surface region. The momentum locations of each spectrum (P1-P15) are indicated by black circles in (c). The spectra are vertically offset for clarity. (b) The corresponding symmetrized EDCs from (a). The vertical ticks mark the peak positions of the symmetrized EDCs from which the superconducting gap is extracted. (c) Schematic Fermi surface of Hg1223 in single Fermi surface region with the corresponding Fermi momentum points (P1-P15) labeled. (d) Angular dependence of the superconducting gap size $\Delta(\theta)$ (filled circles) extracted from (b). The open circles are obtained by flipping the filled circles along 45\(^\circ\) by considering the nodal mirror plane. The angle \(\theta\) is defined as shown in (c). The gap near the nodal region is fitted by a $d$-wave form, $\Delta(\theta) = \Delta_0 \cos(2\theta)$, with $\Delta_0 \approx 41$\,meV (solid curve). (e) Momentum dependence of the superconducting gap plotted as a function of $|\cos(k_x a) - \cos(k_y a)|/2$. The gap near the nodal region is fitted by a linear line, \(\Delta=\Delta_0\cdot |\cos(k_x a) - \cos(k_y a)|/2\), with \(\Delta_0=37\)\,meV (solid line). }
    \label{Fig3_SingleBandGap}
\end{figure}

\begin{figure}[tbp]
    \begin{center}
    \includegraphics[width=1.0\columnwidth,angle=0]{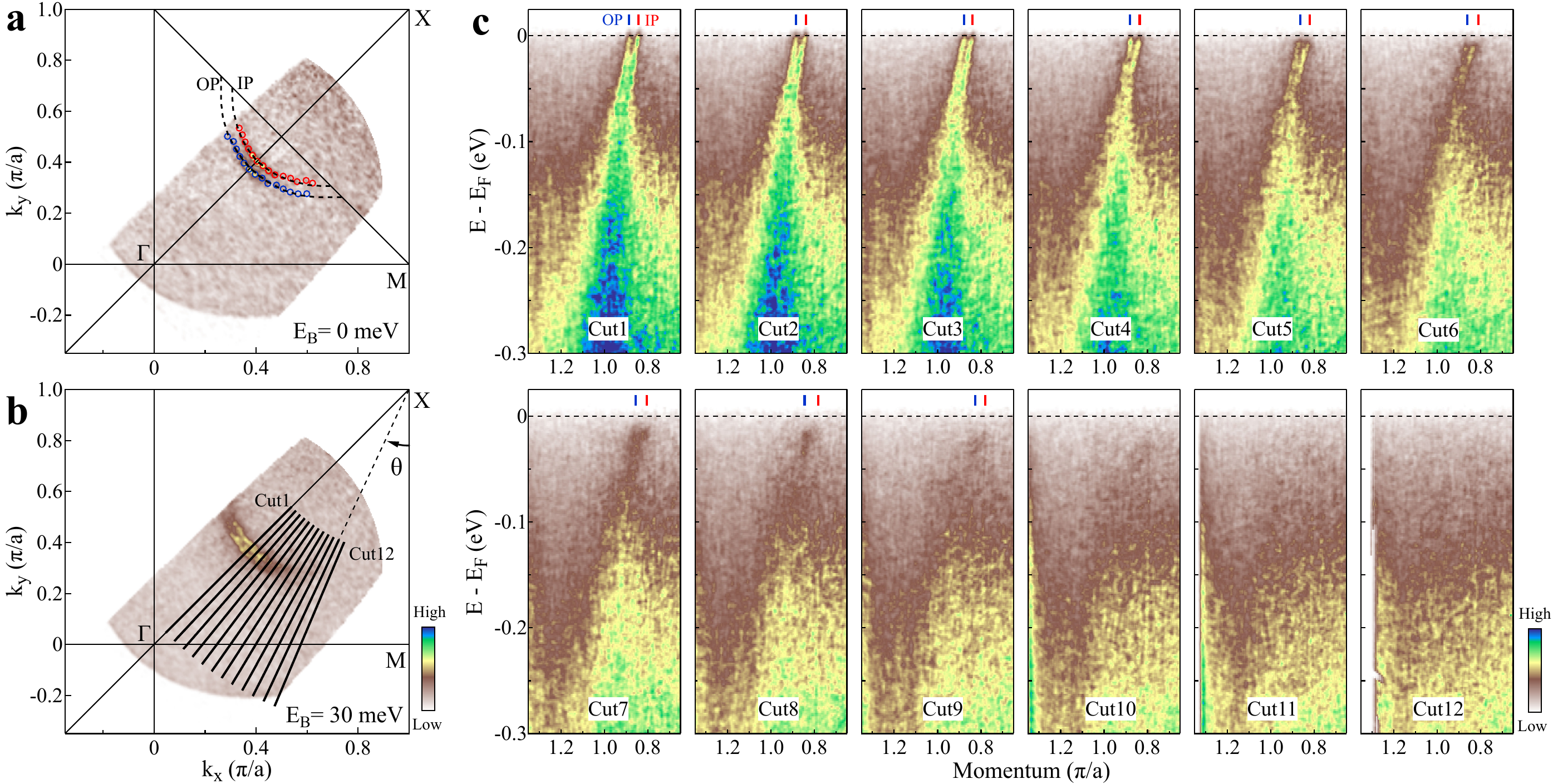}
    \end{center}
    \caption{\textbf{Electronic structure of Hg1223 measured at 30\,K from the double Fermi surface region.} The measurement was performed under a sample bias of -30\,V. (a) Fermi surface mapping of Hg1223 measured at 30\,K in the superconducting state, showing two Fermi surface sheets corresponding to the outer (OP) and the inner (IP) CuO\(_2\) planes. The blue (red) open circles denote the experimentally determined Fermi momenta of OP (IP). The two black dashed curves represent guide to the eyes. (b) The corresponding constant energy contour at a binding energy of 30\,meV. (c) Band structures measured along different momentum cuts. The location of the momentum cuts is shown by black lines in (b). All the momentum cuts cross the \((\pi,\pi)\) point when extrapolated. The blue and red vertical ticks in each panel mark the Fermi momenta for OP and IP, respectively.}
    \label{Fig4_DoubleBandBandStru}
\end{figure}


\begin{figure}[tbp]
    \begin{center}
    \includegraphics[width=1.0\columnwidth,angle=0]{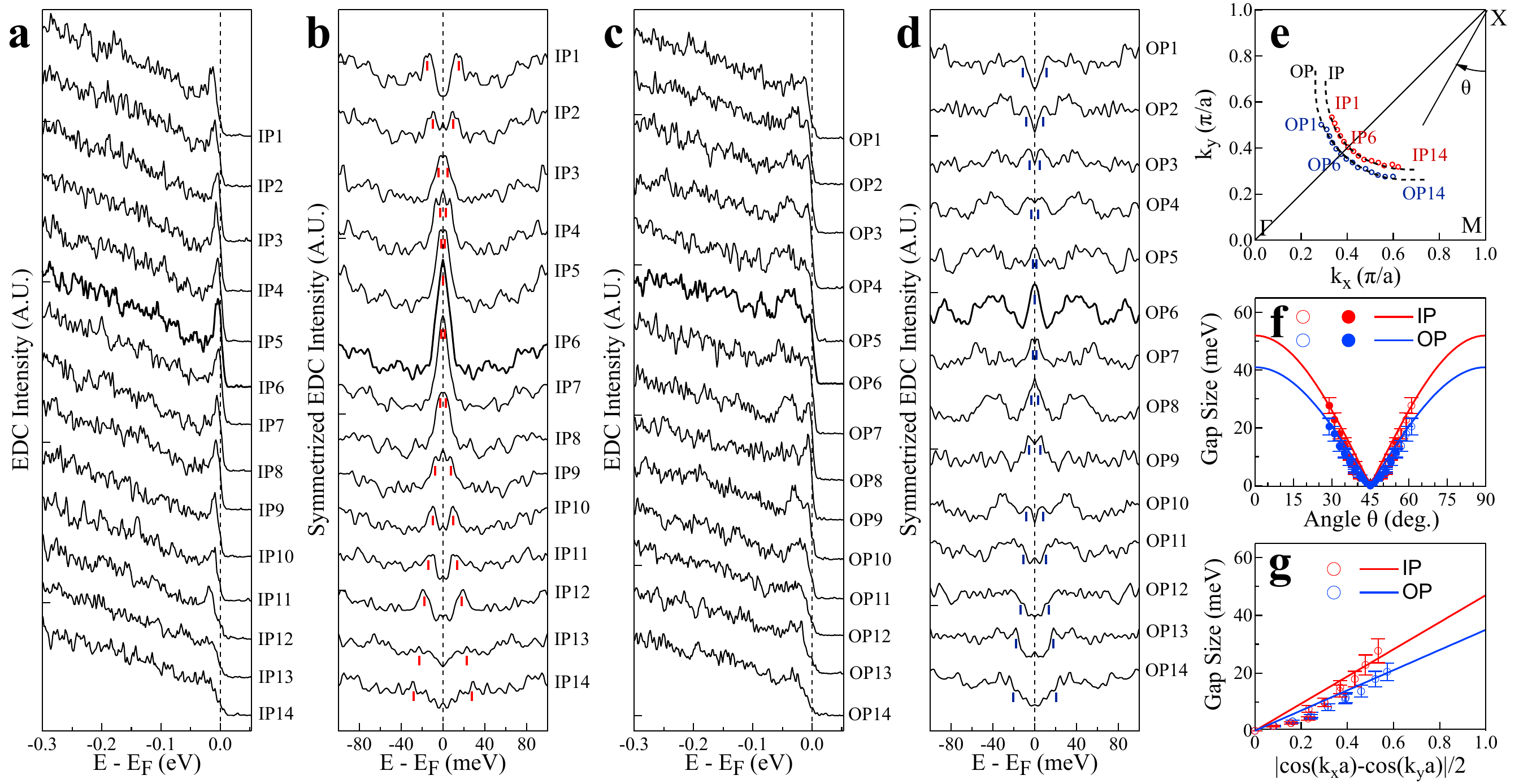}
    \end{center}
    \caption{\textbf{Superconducting gap of Hg1223 measured in the double Fermi surface region at 30\,K.} (a) EDCs measured along the IP Fermi surface. The momentum locations of each spectrum (IP1-IP14) are indicated by red circles in (e). The spectra are vertically offset for clarity. (b) The corresponding symmetrized EDCs from (a). The vertical ticks mark the peak positions of the symmetrized EDCs from which the superconducting gap is extracted. (c) EDCs measured along the OP Fermi surface. The momentum locations of each spectrum (OP1-OP14) are indicated by blue circles in (e). (d) The corresponding symmetrized EDCs from (c). The vertical ticks mark the peak positions of the symmetrized EDCs from which the superconducting gap is extracted. (e) Schematic Fermi surfaces of Hg1223 in double Fermi surface region, showing the inner (IP) and outer (OP) Fermi surface sheets and the corresponding measurement points. (f) Angular dependence of the superconducting gap size $\Delta(\theta)$ for both IP (filled red circles) and OP (filled blue circles) extracted from (b) and (d), respectively. The open circles are obtained by flipping the filled circles along 45\(^\circ\) by considering the nodal mirror plane. The angle \(\theta\) is defined as shown in (e). The gap is fitted by a $d$-wave form, $\Delta(\theta) = \Delta_0 \cos(2\theta)$, with $\Delta_0 \approx 52$\,meV for IP (red solid curve) and $\Delta_0 \approx 41$\,meV for OP (blue solid curve). (g) Momentum dependence of the superconducting gap plotted as a function of $|\cos(k_x a) - \cos(k_y a)|/2$. The gap is fitted by a linear line, \(\Delta=\Delta_0\cdot |\cos(k_x a) - \cos(k_y a)|/2\), with \(\Delta_0=47\)\,meV for IP (red solid line) and \(\Delta_0=35\)\,meV for OP (blue solid line).}
    \label{Fig5_DoubleBandGap}
\end{figure}

\end{document}